\documentclass[amsmath,amssymb,10pt,aps,notitlepage,twocolumn]{revtex4-1} 
 \usepackage[colorlinks=true]{hyperref}
 \usepackage{lipsum}
 \usepackage{graphicx}
 \usepackage{textcomp}
\usepackage[dvipsnames]{xcolor}
\usepackage{latexsym}
\usepackage{amsmath}
\usepackage{amsthm}
\usepackage{amssymb}
\usepackage{epstopdf} 
\usepackage{enumerate}
\usepackage{setspace} % controllabel line spacing
\usepackage{dcolumn}% Align table columns on decimal point
\usepackage{bm}% bold math
\usepackage{setspace} % controllabel line spacing
\usepackage{slashed}
\usepackage{color}
\usepackage{xcolor}
\usepackage{youngtab}
\usepackage{tikz}
\usepackage{tikz-3dplot}
\usepackage{braket}
\usetikzlibrary{shapes,snakes,arrows,chains,matrix,positioning,scopes,calc}
\usetikzlibrary{decorations.markings}

\usepackage[tabbotcap]{subfigure}

\allowdisplaybreaks
 
\begin{document}
\title{Non-Fermi liquid induced by Bose metal with protected subsystem symmetries}

\author{SangEun Han}
\author{Yong Baek Kim}

\affiliation{Department of Physics, University of Toronto, Toronto, Ontario M5S 1A7, Canada
}

\date{\today}   

\begin{abstract}

Understanding non-Fermi liquids in dimensions higher than one, has been a subject of great interest. Such phases may serve as parent states for other unconventional phases of quantum matter, in a similar manner that conventional broken symmetry states can be understood as instabilities of the Fermi liquid. In this work, we investigate the emergence of a novel non-Fermi liquid in two dimensions, where the fermions with quadratic band-touching dispersion interact with the Bose metal. The bosonic excitations in the Bose metal possess an extended nodal-line spectrum in momentum space, which arises due to the subsystem symmetry or the restricted motion of bosons. 
Using renormalization group analysis and direct computations, we show that the extended infrared (IR) singularity of the Bose metal leads to a line of interacting fixed points of novel non-Fermi liquids, where the anomalous dimension of the fermions varies continuously, akin to the Luttinger liquid in one dimension. Further, the generalization of the model with multiple low-energy excitations is used to explore other unusual features of the resulting ground state.

\end{abstract}

\maketitle

\emph{Introduction--}
Classification of gapless quantum ground states of interacting fermions is an outstanding question in modern theory of quantum matter. Deciphering the origin and instability patterns of such phases holds the key for understanding quantum critical phases and novel broken symmetry or topological phases that may arise thereof \cite{Sachdev475,RevModPhys.79.1015,WitczakKrempa2014}. In one dimension, the Luttinger liquid \cite{Haldane_1981} is a well-known example going beyond the paradigm of the Fermi liquid that is the standard model of conventional metals. It does not have well-defined quasiparticles and is characterized by a continuously-varying exponent that describes the algebraic correlations in space-time \cite{Haldane_1981}. In higher dimensions, there have been numerous studies of non-Fermi liquids that may arise from the long-range interaction between gapless fermions and critical bosonic excitations \cite{PhysRevB.50.14048,PhysRevB.50.17917,PhysRevB.50.17917,Polchinski1994,PhysRevB.46.5621,PhysRevB.64.195109,PhysRevB.78.085129,PhysRevB.89.165114,PhysRevLett.84.5608,PhysRevLett.111.206401,PhysRevLett.122.187601,Nayak1994,Nayak19942,PhysRevB.80.165102,PhysRevB.82.075127,PhysRevB.82.075128,PhysRevB.82.045121,PhysRevB.88.245106,doi:10.1146/annurev-conmatphys-031016-025531}. Nonetheless, these systems are often in the strong coupling limit, which makes it hard to find a controlled theoretical framework \cite{PhysRevB.80.165102,PhysRevB.82.075127,PhysRevB.82.075128,PhysRevB.82.045121,PhysRevB.88.245106,doi:10.1146/annurev-conmatphys-031016-025531}.

Considering the fermion-boson interactions in two and three dimensions, the bosons would condense at zero temperature if they are gapless at a specific momentum (unless they are Goldstone modes or gauge fields). This is the reason why such interactions are mostly studied near a quantum critical point for the bosons, where they remain gapless down to zero temperature. If the interacting boson systems remain gapless at zero temperature and exist as critical phases, such systems, when they are coupled to fermions, may offer a novel platform (i.e.~without quantum critical point) for possible emergence of non-Fermi liquid states. Such bosonic ground states may be called the Bose metals. These phases have recently gotten much attention due to the connection to the physics of fracton quantum order \cite{PhysRevLett.94.040402,PhysRevA.83.042330,PhysRevB.92.235136,PhysRevB.95.115139,PhysRevB.74.224433,PhysRevB.96.195139,PhysRevX.8.031051,Nandkishore2019,Pretko2020,you2020fracton,PhysRevResearch.2.013162,10.21468/SciPostPhys.10.2.027,10.21468/SciPostPhys.10.1.003,PhysRevB.103.245128}. For example, the Bose metal phases that arise from the ring-exchange interactions are protected by the sub-system symmetry, i.e.~the boson number for each row and column of the underlying lattice is separately conserved \cite{PhysRevB.66.054526,10.21468/SciPostPhys.10.2.027}. Here, the elementary excitations are gapless in an extended region in the momentum space, which prohibits the condensation \cite{PhysRevB.66.054526,Sachdev2002,PhysRevB.100.024519,PhysRevB.104.014517}. A crucial question in the context of possible non-Fermi liquid states is whether the sub-system symmetries mentioned above can be protected even in the presence of the coupling to the fermions as such symmetries are the keys to the existence of the critical phase of bosons from the first place.  
 
\begin{figure}[b]
\subfigure[]{
\includegraphics[scale=0.8]{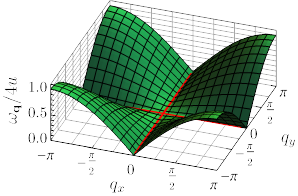}\label{fig:bosemetal}
}\subfigure[]{
\includegraphics[scale=0.8]{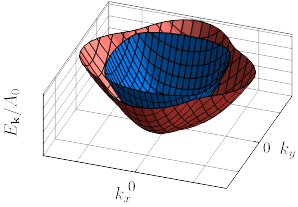}\label{fig:2band}
}
\caption{Plots of the dispersions in the Bose metal and two-band model of fermions with $p_{x}$ and $p_{y}$ orbitals. (a) The Bose metal has nodal lines at $k_{x}=0$ and $k_{y}=0$ (red lines). (b) The dispersion of the fermions with $A_{x}/A_{0}=0.1$ and $A_{z}/A_{0}=0.4$. The blue and red bands are the top and bottom bands touching at $(k_x, k_y) = (0,0)$.
}
\end{figure}

In this work, we show that non-Fermi liquids arise when the fermions with a quadratic dispersion interact with the Bose metal in two dimensions. Using the renormalization group, we explain how the sub-system symmetries of the Bose metal can be still protected even in the presence of the fermions. In particular, we consider the ring-exchange model of the Bose metal on the square lattice \cite{PhysRevB.66.054526}, the dispersion of the low energy excitations are given by $\omega_q \sim |{\sin (q_x/2)} {\sin (q_y/2)}|$ (Fig.~\ref{fig:bosemetal}). The IR singularity along the nodal lines at $q_{x}=0$ an $q_{y}=0$ may provide a seed for non-Fermi liquids when the Bose metal is coupled to fermions.

\begin{figure}
\subfigure[]{
\includegraphics[height=12.5em]{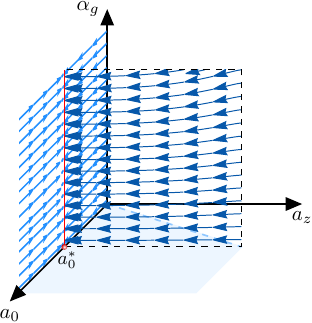}\label{fig:alpha_g}}\subfigure[]{
\includegraphics[height=12.5em]{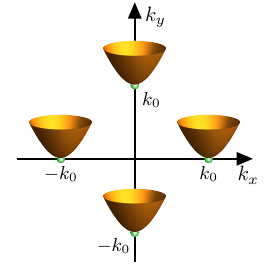}\label{fig:multi}
}
\caption{(a) RG flow diagram in terms of $a_{0},a_{z},\alpha_{g}$ for $a_{0}=a_{0}^{*}$ and $a_{z}=0$ slices. Here, the initial value $a_{x,\text{init}}=0$ is used.
The red solid line represents the line of fixed points, $(a_{0}^{*},a_{x}^{*},a_{z}^{*},\alpha^*_g)=(1.4518,0,0,\alpha^*_g)$, where $\alpha_{g}^{*}$ depends on the initial values of other parameters.
(b) The location of the fermion band-touching points, in momentum space, in the model with multiple low-energy excitations. 
The locations of band-touching points are $\mathbf{k}=(\pm k_{0},0)$ and $(0,\pm k_{0})$.
}
\end{figure}

We first consider the interaction between the fermions with a quadratic dispersion at the Brillouin zone center and the Bose metal. Focusing on the low-energy continuum limit of the bosonic excitations, we show that the one-loop fermion self-energy acquires the logarithmic correction, ${\rm Re} \Sigma \sim \omega \ln \omega$, while the bosonic dispersion is not renormalized. Motivated by this discovery, we then consider the continuum limit of the two band model of fermions with $p_x, p_y$ orbitals and the quadratic band-touching (Fig.~\ref{fig:2band}), which interact with the Bose metal. Using the renormalization group analysis, we find that a line of interacting fixed points arises (see Fig.~\ref{fig:alpha_g}), where the fermions acquire an anomalous dimension that varies continuously with the fixed point values of the dimensionless interaction parameter. In addition, the Bose metal is not renormalized or the subsystem symmetry is preserved at the fixed point. The continuously-varying anomalous dimension is akin to the Luttinger liquid \cite{Haldane_1981}. 

In order to further explore the influence of the IR singularity from the nodal lines, we also consider the quadratic-band-touching fermions located at four different symmetric positions in the Brillouin zone, namely $\mathbf{k}=(\pm k_{0},0),(0,\pm k_{0})$ (see Fig.~\ref{fig:multi}). The lattice boson dispersion, $\omega_q \sim |{\sin (q_x/2)} {\sin (q_y/2)}|$, allows two kinds of IR singularities. The first kind acts mainly within the same small area in momentum space around each quadratic band-touching point. This involves the small momentum transfer ${\bf q} = (\delta q_x, \delta q_y)$, where $|\delta q_x|, |\delta q_y| \ll k_0$, with the relevant bosonic excitations described by $\omega_q \sim |\delta q_x \delta q_y|$. The second kind involves the large momentum transfer (in one of the two directions) between the reflection-related areas in momentum space, namely ${\bf q} = (\delta q_x, 2k_0)$ or ${\bf q} = (2k_0, \delta q_y)$, with the corresponding bosonic soft mode $\omega_q \sim |\delta q_x \sin (k_0)|$ or $\omega_q \sim |\sin (k_0) \delta q_y|$. Considering the one-loop self-energy correction, it is found that the small momentum process in each area dominates the low energy scaling. Hence the same low energy fixed point found earlier should apply in the scaling limit. On the other hand, the large momentum transfer in one of the momentum directions leads to a non-analytic correction to the fermion self-energy. For example, at the one-loop level, we find ${\rm Re} \Sigma \sim \omega \ln \omega + \omega^{3/2}$ and ${\rm Im} \Sigma \sim \omega + \omega^{3/2}$. Here the subleading $\omega^{3/2}$ is from the large momentum transfer between different areas in momentum space. Hence the second derivative $\partial^2 {\rm Im} \Sigma / \partial \omega^2 \sim \omega^{-1/2}$ diverges, which is in principle visible in the fermion scattering rate measurement in ARPES \cite{RevModPhys.75.473}. In this sense, the large momentum transfer process between different areas in momentum space is dangerously irrelevant.
and may play an important role in the characterization of the underlying non-Fermi liquid ground state.

\emph{Bose metal--}
We first consider the ring-exchange model of bosons on the square lattice \cite{PhysRevB.66.054526},
\begin{equation}
H = \sum_{\bf r} \left [ {U \over 2} 
\left (n_{\bf r} - \bar{n}
 \right )^2 - K \cos (\Delta_{xy} \phi_{\bf r}) \right ]  \ ,
\end{equation}
where $\Delta_{xy} \phi_{\bf r} = \phi_{\bf r} - \phi_{{\bf r}+{\hat {\bf x}}} - \phi_{{\bf r}+{\hat {\bf y}}} + \phi_{{\bf r}+{\hat {\bf x}} + {\hat {\bf y}}}$ is defined on each plaquette. Here $n_{\bf r}$ and $\phi_{\bf r}$ correspond to the boson number and the phase of the boson wave function at each lattice site ${\bf r}=(x,y)$, which satisfy the canonical commutation relation, $[\phi_{\bf r}, n_{\bf r'}] = i \delta_{{\bf r}{\bf r'}}$. 
$\bar{n}$ is the mean boson density.
The phase $\phi_{\bf r}$ is $2\pi$ periodic, $\phi_{\bf r} = \phi_{\bf r} + 2\pi$, so that $n_{\bf r}$ takes the integer values. The ring-exchange interaction correspond to the two-particle correlated hopping in each plaquette such that the boson number in each row and column of the lattice is preserved. 

It was shown that the Bose metal phase is a stable ground state of this model for a range of parameters \cite{PhysRevB.66.054526,10.21468/SciPostPhys.10.2.027}, where the effective low energy action can be obtained via the expansion,
$\cos (\Delta_{xy} \phi) \sim 1 - {1 \over 2} (\Delta_{xy} \phi)^2$, 
\begin{align} 
\mathcal{S}=&\frac{1}{2}\int\frac{d^{2}q}{(2\pi)^{2}}\int_{-\infty}^{\infty}\frac{d\omega_{n}}{2\pi}(\omega_{n}^{2}+\omega_{\mathbf{q}}^{2})|\phi(\omega_{n},\mathbf{q})|^{2} \ ,
\end{align}
where $\omega_{\mathbf{q}}\equiv 4u|\sin(q_{x}/2)\sin(q_{y}/2)|$ for $q_{x},q_{y}\in(-\pi,\pi)$.
Here we set $U=1$ and $u = \sqrt{U K}$. The nodal lines at $q_x=0, q_y=0$ reflect the presence of infinite number of conserved quantities (Fig.~\ref{fig:bosemetal}).
The corresponding symmetries involve the invariance of the action under $\phi (x,y) \rightarrow \phi (x,y) + \Phi_x (x) + \Phi_y (y)$, where
$\Phi_x (x), \Phi_y (y)$ are arbitrary functions of $x$ and $y$ \cite{PhysRevB.66.054526,10.21468/SciPostPhys.10.2.027}. 
The lattice action for the Bose metal is symmetric under $C_{4}$ rotation, inversion, time reversal, and reflection about $x,y$ axes. 
The continuum limit is rather subtle and $\Delta_{xy} \phi \rightarrow \partial_x \partial_y \phi$ is well defined while $\partial_x \phi$ and $\partial_y \phi$ are not \cite{10.21468/SciPostPhys.10.2.027}. 
The resulting continuum action, ${\cal S}  \sim \int d\tau \int d^2 x [ (\partial_\tau \phi)^2 + u^2 (\partial_x \partial_y \phi)^2 ]$ is
also consistent with $\omega_{\bf q} = u |q_x q_y|$ for small momentum $q_x, q_y$.
The symmetries mentioned above imply that the coupling to fermion bilinears, $\psi^{\dagger} M \psi$, where $M$ is a symmety-allowed matrix representation, involves the form factor, $\mathcal{F} ({\bf q}) = 4\sin (q_x/2) \sin (q_y/2)$, in the lattice model. Thus, the interaction vertex has the form $\sum_{{\bf k},{\bf q}} \psi^{\dagger}_{{\bf k}+{\bf q}} M \psi_{\bf k} \mathcal{F} ({\bf q}) \phi_{\bf q}$, which leads to $\psi^{\dagger} M \psi (\partial_x \partial_y \phi)$ in the continuum limit  \cite{SM}.

\emph{p-wave orbital model and preliminary analysis--}
Let us consider the fermions in $p_x, p_y$ orbitals on the two-dimensional square lattice.
In the continuum limit, the model can be written as
\begin{eqnarray}
H&=&\sum_{\mathbf{k}}\Psi^{\dagger}_{\mathbf{k}}\mathcal{H}(\mathbf{k})\Psi_{\mathbf{k}}, \cr
\mathcal{H}(\mathbf{k})&=&\;A_{0}k^{2}\sigma_{0}+A_{x}(2k_{x}k_{y})\sigma_{x}+A_{z}(k_{x}^{2}-k_{y}^{2})\sigma_{z},\label{eq:fermi_Hamil}
\end{eqnarray}
where $\Psi^{\intercal}=(c_{x}, c_{y})$ is a two-component spinor, $c_{x,y}$ is the fermion annihilation operator for $p_{x,y}$ orbital, $\sigma_{i}$ is the Pauli matrix, and $\sigma_{0}$ is the $2\times2$ identity matrix. The corresponding fermion dispersion is $E(\mathbf{k})=A_{0}k^{2}\pm\sqrt{4A_{x}^{2}k_{x}^{2}k_{y}^{2}+A_{z}^{2}(k_{x}^{2}-k_{y}^{2})^{2}}$. We assume $A_{0,x,z}$ are positive. We focus on the case of $A_{0}>\text{max}(A_{x},A_{z})$, in which two bands are quadratically touching at $\mathbf{k}=(0,0)$ (Fig.~\ref{fig:2band}). The Hamiltonian is invariant under $C_{4}$ rotation, $\mathcal{U}_{C_{4}}\Psi=i\sigma_{y}\Psi$, and the reflection about $x$ and $y$ axes, $\mathcal{U}_{\mathcal{R}_{x,y}}=\pm\sigma_{z}\Psi$.

When these fermions interact with the bosonic excitations in the Bose metal, the action for the  interacting model can be written as 
\newpage
\begin{align}
\mathcal{S} =& \mathcal{S}_0 + \mathcal{S}_{\text{int}},\notag\\
\mathcal{S}_{0}=&\int_{x,\tau}\Psi^{\dagger}(\partial_{\tau}+\mathcal{H})\Psi+\frac{1}{2}[(\partial_{\tau}\phi)^{2}+u^{2}(\partial_{x}\partial_{y}\phi)^{2}] ,\notag\\
\mathcal{S}_{\text{int}}=&g\int_{x,\tau}(\partial_{x}\partial_{y}\phi)(\Psi^{\dagger}\sigma_{x}\Psi),\label{eq:int} 
\end{align}
where $\phi$ is the phase field of the bosons, $\Psi$ and $\mathcal{H}$ are the fermion fields and their Hamiltonian introduced earlier. 
The tree-level scaling dimensions of fields and parameters are $[\Psi]=d/2$, $[\phi]=(d-z)/2$, $[A_{i}]=[u]=z-2$. 
In particular, the scaling dimension of $g$ is $[g]=(3z-d-4)/2$. Hence, for $d=2$ and $z=2$, this interaction is marginal, $[g]=0$.
Before going further, we introduce the dimensionless parameters, $\alpha_{g}\equiv g^{2}/\pi^{2}u^{3}$, and $a_{i}=A_{i}/u$ where $i=0,x,z$.
Below, we use the fermion and boson propagators given by
$G(i\omega_{n},\mathbf{k})=(-i\omega_{n}\sigma_{0}+\mathcal{H})^{-1}$ and $D(i\omega_{n},\mathbf{q})=(\omega_{n}^{2}+u^{2}q_{x}^{2}q_{y}^{2})^{-1}$, respectively.

We first consider the model with $A_{0} \not=0$, but all other $A_x, A_z = 0$.
The one-loop boson self-energy is
\begin{align}
\Pi&(i\omega_{n},\mathbf{q})=-g^{2}q_{x}^{2}q_{y}^{2}\int_{\Lambda}\frac{d^{2}p}{(2\pi)^{2}}\int_{-\infty}^{\infty}\frac{d\Omega_{m}}{2\pi} \notag\\
&\quad \times \text{Tr}[\sigma_{x}G(i\Omega_{m}+i\omega_{n},\mathbf{p}+\mathbf{q})\sigma_{x}G(i\Omega_{m},\mathbf{p})],
\end{align}
where $\Lambda$ is the UV cutoff. 
In our setup, since the quasiparticle poles of the fermions are in the same half-plane, the dynamics of the bosons do not get  renormalized for small $\alpha_{g}$ and small external momenta; in this regime, perturbative calculations are valid. 
Hence, in the low-energy limit, the dynamics of the boson are not changed and the bosonic part of the action is the same as the original Bose metal.
The one-loop fermion self-energy is given by
\begin{align}
\Sigma(i\omega_{n})
=&\;g^{2}\int_{\Lambda}\sigma_{x}G(i\omega_{n}+i\Omega_{m},\mathbf{p})\sigma_{x}D(i\Omega_{m},\mathbf{p})[\mathcal{F}(\mathbf{p})]^{2},\label{eq:orig_area}
\end{align}
where $\int_{\Lambda}=\int_{\Lambda}\frac{d^{2}p}{(2\pi)^{2}}\int_{-u\Lambda^{2}}^{u\Lambda^{2}}\frac{d\Omega_{m}}{2\pi}$.
When $|\omega_{n}|/u\Lambda^{2}\ll1$, after analytic continuation, $i\omega_{n}\rightarrow\omega+i\eta$, Eq.~\ref{eq:orig_area} leads to 
the following fermion self-energy in real frequency.
\begin{align}
\Sigma(\omega) \approx \alpha_{g} \left [ C_{\mathcal{I},1} \ i \omega
+C_{\mathcal{R},\text{log}} \ \omega \ln \left ( \frac{\omega}{u\Lambda^{2}} \right ) \right ] \sigma_{0} \ ,
\end{align}
where $C_{i}$'s are constants depending on $a_{0}$, and we assume $\omega>0$ for simplicity. 
The logarithmic correction encourages the renormalization group analysis, which we present below.

\emph{Renormalization group analysis--} Based on the above observation, we perform the momentum-shell renormalization group analysis for the full model Hamiltonian.
The renormalization of the boson self-energy $\delta\Pi$ at the one-loop level can be shown to vanish, $\delta\Pi=0$, when $A_{0}>\text{max}(A_{1},A_{3})$. 
Thus the bosonic action is not renormalized, just like the previous example with $A_0 \not = 0$ and $A_x, A_z = 0$.

The renormalization correction to the fermion self-energy is given by
\begin{align}
\delta \Sigma (i\omega_{n},\mathbf{k})
\approx&
-\alpha_{g}\ell[ F_{\omega}(-i\omega_{n})\sigma_{0}+F_{0}A_{0}k^{2}\sigma_{0}\notag \\
&+F_{x}(2A_{x}k_{x}k_{y})\sigma_{x}+F_{z}A_{z}(k_{x}^{2}-k_{y}^{2})
],\label{eq:deltaSigma}
\end{align}
where $\ell=\ln (\Lambda/\mu)$, $\Lambda$ and $\mu$ are UV and IR cutoffs \footnote{During coarse-graining, we integrate out the degrees of freedom inside the momentum shell between $\Lambda e^{-\ell}<k<\Lambda$. It includes $k_{x}=0$ and $k_{y}=0$ momenta which are related to the low-energy degrees of freedom of bosons. In our model, the interaction vertex, Eq.~\ref{eq:int}, includes the momentum dependent form factor, $k_{x}k_{y}$. 
Hence, the contribution from the degrees of freedom that are related to $k_{x}=0$ or $k_{y}=0$ momenta is highly suppressed by the form factor of the interaction vertex. Therefore, integrating out the low-energy degree of freedoms of bosons for $k_{x}=0$ or $k_{y}=0$ does not lead to any singular behavior.}.
Here $F_{i}$'s are functions of $a_{0}$, $a_{x}$, and $a_{z}$, and their definitions are provided in Supplementary Materials \cite{SM}. 
Note that $F_{\omega,0}$ are positive, but $F_{z}$ is negative for $a_{0}>a_{x,z}$.
The vertex correction is found as $\delta \Gamma_{g}=\alpha_{g}F_{g}\ell$ where $F_{g}$ is a function of $a_{0}$, $a_{x}$, $a_{z}$, 
and positive for $a_{0}>a_{x,z}$. Moreover, $F_{\omega}-F_{g}\geq0$ for $a_{0}>a_{x,z}$  \cite{SM}.

In terms of dimensionless parameters $a_{i}$ and $\alpha_{g}$, the RG equations up to one-loop order are given by
\begin{align}
\frac{1}{a_{i}}\frac{da_{i}}{d\ell}=&-\alpha_{g}(F_{\omega}-F_{i}),\;\;
\frac{1}{\alpha_{g}}\frac{d\alpha_{g}}{d\ell}=-2\alpha_{g}(F_{\omega}-F_{g}).
\end{align}
We now analyze the above RG equations.
In the first place, $da_{z}/d\ell$ is negative because $F_{\omega}$ is positive and $F_{z}$ is negative. Thus, $a_{z}$ decreases under the RG flow and may approach $a_{z}\rightarrow0$.
With $a_{z}\rightarrow0$, we find fixed point values of $a_{0}$ and $a_{x}$, $(a_{0}^{*},a_{x}^{*})=(1.4518,0)$.\
On the other hand, $\alpha_{g}$ decreases from an initial value as $a_{z}$ diminishes in the RG flow because $F_{\omega}-F_{g}$ is positive.
As $a_{z} \rightarrow 0$, $F_{\omega}$ and $F_{g}$ approach the same value. It is a consequence of the Ward identity, $\lim\limits_{i\omega_{n}\rightarrow0}\text{Tr}[\partial\delta\Sigma/\partial(i\omega_{n})]/2=\delta\Gamma_{g}$, because the interaction vertex (Eq.~\ref{eq:int}) commutes with the fermion Hamiltonian (Eq.~\ref{eq:fermi_Hamil}) when $a_{z}\rightarrow0$. This, combined with the fact that the bosonic action is not renormalized, makes $\alpha_{g}$ to approach a finite value under the RG flow as $a_{z}\rightarrow0$. The fixed point value of $\alpha_g$, however, depends on the initial values of other dimensionless parameters, as shown in Fig.~\ref{fig:alpha_g}.
This means there exists a line of stable fixed points, $(a_{0}^{*},a_{x}^{*},a_{z}^{*},\alpha_{g}^{*})=(1.4518,0,0,\alpha_{g}^{*})$, characterized by continuously varying fixed point values of $\alpha^*_g$. Some numerical solutions of the RG equations are shown in Fig.~S1 in Supplementary Materials \cite{SM}.

Along the line of stable fixed points, the anomalous dimension of the fermions is finite as $\alpha_{g} \not= 0$ and is given by 
$\eta_{f}=\alpha_{g}C_{f}$,
where $C_{f}=F_{\omega}(a_{0}=1.4518,a_{x}=a_{z}=0)=0.07438$.
This means the fermion propagator behaves as $G_{F}(i\omega_{n},\mathbf{k})\sim 1/(-i\omega+E_{\mathbf{k}})^{1-\eta_{f}}$ in the low-energy limit, 
and there is no quasiparticle pole.
The absence of a quasiparticle pole is the primary signature of non-Fermi liquid behavior \cite{doi:10.1146/annurev-conmatphys-031016-025531}.
Moreover, the anomalous dimension of the fermions changes continuously as $\alpha_{g}$ is varied along the line of stable fixed points.
This behavior is reminiscent of the Luttinger liquids in one dimension \cite{Haldane_1981}. In this sense, we may consider the line of fixed points 
as a two-dimensional version of the Luttinger liquids.

\emph{The model with multiple low-energy excitations--}
In order to further examine the influence of the nodal lines in the Bose metal, we consider multiple fermions 
at symmetry-related locations in the momentum space. 
Let us assume that the fermions with the quadratic dispersion, $A_{0}k^{2}$, are located at ${\bf k} = (\pm k_0, 0), (0, \pm k_0)$ (Fig.~\ref{fig:multi}).
We first consider small areas in momentum space near each band-touching point.
For small momentum transfer in both $k_x$ and $k_y$ directions, we only need to consider the scattering processes within the same area.
However, when the momentum transfer is large in one of the directions and small in the other, two different areas related by the reflection about $k_x$ or $k_y$ axis,
can be connected by such a momentum transfer in the low energy limit. 

To illustrate this point, let us focus on the area near $\mathbf{k}=(k_{0},0)$ and compute the fermion self-energy.
In the case of the small momentum transfer, $\mathbf{q}=(\delta q_x, \delta q_y)$, the dispersion and form factor for the bosonic excitations are 
given by $\omega_{\mathbf{q}}=u|\delta q_{x}\delta q_{y}|$ and $\mathcal{F}(\mathbf{q})=\delta q_{x}\delta q_{y}$. 
The one-loop fermion self-energy due to the small momentum transfer within the same area has the same form as Eq.~\ref{eq:orig_area}, and after analytic continuation (here and in the following, we assume $\omega>0$ after analytic continuation for simplicity),
\begin{equation}
\Sigma_{\text{small}}(\omega) \approx \alpha_{g}  [ C_{\mathcal{R},1}^{\text{small}} \ \omega + C_{\mathcal{R},\text{log}}^{\text{small}} \ \omega \ln \left ( \frac{\omega}{u\Lambda^{2}} \right ) +  C_{\mathcal{I},1}^{\text{small}} \ i\omega  ].
\label{eq:small}
\end{equation}
Now let us consider the large momentum transfer between different areas near ${\bf k} = (k_0, 0)$ and ${\bf k} = (-k_0, 0)$.
The corresponding momentum transfer between these areas for low energy bosonic excitations can be written as $\mathbf{q}=(-2k_{0}, \delta q_y)$. 
The dispersion and form factor are given by $\omega_{\bf q}=2u| (\delta q_{y}) \sin k_{0}|$ 
and $\mathcal{F}({\bf q})=2 (\delta q_{y}) \sin k_{0}$.
Note that we do not consider the large momentum transfer between different areas located near $\mathbf{k}=(0,\pm k_{0})$ and $(k_{0},0)$ 
as it is not a low energy process.
Now the one-loop fermion self-energy at ${\bf k}=(k_0,0)$ due to the momentum transfer $\mathbf{q}=(-2k_{0}, \delta q_y)$ is given by
\begin{align}
&\Sigma_{\text{large}}(i\omega_{n})
\approx \alpha_g\Big(\mathcal{C}_{\mathcal{R},3/2}^{\text{large}} {\omega_{n}^{3/2} \over (u \Lambda^2)^{1/2}}
+\mathcal{C}_{\mathcal{I},1}^{\text{large}} i\omega_{n} 
\notag\\
&\quad\quad\quad\quad\quad\quad\quad\quad+ \mathcal{C}_{\mathcal{I},3/2}^{\text{large}} {i\text{sgn}(\omega_{n})|\omega_{n}|^{3/2} \over (u \Lambda^2)^{1/2}} \Big).\label{eq:large}
\end{align}
The detail is shown in the Supplementary Materials \cite{SM}.

After analytic continuation, the total fermion self-energy in real frequency, $\Sigma_{\text{tot}}(\omega)\equiv\Sigma_{\text{small}}(\omega)+\Sigma_{\text{large}}(\omega)$, is obtained.
The real part of this fermion self-energy is 
\begin{align}
\text{Re} \Sigma_{\text{tot}}(\omega) =& \alpha_g  \left [ \ C_{\mathcal{R},1}^{\text{tot}}  \omega  + C_{\mathcal{R},\text{log}}^{\text{tot}}  \omega  \ln\Big(\frac{\omega}{u\Lambda^{2}}\Big)\notag \right . \\
&\quad\quad \left . + C_{\mathcal{R},3/2}^{\text{tot}} {\omega^{3/2} \over (u \Lambda^2)^{1/2}}\right ].
\end{align}
The leading contribution, $\omega\ln\omega$, comes from the small momentum transfer process within the same area. 
It implies that the same renormalization group fixed point obtained earlier for the model with a single low-energy mode would describe the low energy properties of this non-Fermi liquid state.

On the other hand, the imaginary part of the total fermion self-energy is given by
\begin{align}
\text{Im} \Sigma_{\text{tot}}(\omega) = 
\alpha_g \left [ C_{\mathcal{I},1}^{\text{tot}} \ \omega + C_{\mathcal{I},3/2}^{\text{tot}} { \omega ^{3/2} \over (u \Lambda^2)^{1/2}} \right ] .
\end{align}
While the leading behavior is linear in $\omega$, the non-analytic correction $\omega^{3/2}$ comes from the large momentum process.
Note that the second derivative, ${\partial^{2}\text{Im} \Sigma(\omega) }/{\partial\omega^{2}}\propto\omega^{-1/2}$, is singular
in the low energy limit. 
If we only kept the leading order contribution from the small momentum transfer within the same area, we would not be able to
find such a singular behavior. 
In this sense, the large momentum transfer process is dangerously irrelevant. 
It should also be noted that both the small and large momentum processes do not renormalize the bosonic action of the Bose metal \cite{SM}.

\emph{Discussion--}
In the two-dimensional Bose metal phase, one can define the currents, $J_{0} = \partial_{\tau} \phi$, $J_{xy} = u^{2}( \partial_{x}\partial_{y}\phi)$, and the continuity equation, $\partial_{\tau} J_{0} = \partial_{x}\partial_{y}J_{xy}$, can be derived from the equation of motion, $\partial_{\tau}^{2}\phi=\partial_{x}\partial_{y}(u^{2}\partial_{x}\partial_{y}\phi)$ \cite{10.21468/SciPostPhys.10.2.027,10.21468/SciPostPhys.10.1.003,PhysRevB.103.245128}. Noting that the conjugate momentum $\pi_{0} = \partial_{\tau} \phi \sim n (x,y,\tau)$ is the boson number, one can obtain the conserved quantities, $Q_{x}(\tau) = \int dy J_{0}(x,y,\tau)$, $Q_{y}(\tau) = \int dx J_{0}(x,y,\tau)$, which are related to the total number of bosons on each row and column of the lattice. Since $\partial Q_x / \partial \tau = \int dy  \partial_x \partial_y J_{xy}$, $\partial Q_y / \partial \tau = \int dx  \partial_x \partial_y J_{xy}$, and $J_{xy}$ is well defined in the continuum limit \cite{10.21468/SciPostPhys.10.2.027}, we obtain $\partial Q_x / \partial \tau = \partial Q_y / \partial \tau = 0$, with the appropriate boundary condition, and hence $Q_x$, $Q_y$ are conserved.

In the presence of the interaction with fermions, we showed that the non-Fermi %(or non-Fermi for short)
 liquid fixed point exists, where the bosonic action is not renormalized as long as $A_{0}>\text{max}(A_{x},A_{z})$ in the microscopic model. Therefore, at the fixed point $g = g^*$, the continuity equation is simply modified to $\partial_{\tau} J_{0} = \partial_{x}\partial_{y} {\widetilde  J}_{xy}$, where ${\widetilde J}_{xy} = J_{xy} + g^* \partial_x \partial_y (\Psi^{\dagger}\sigma_{x}\Psi)$.
This leads to $\partial Q_x / \partial \tau = \int dy  \partial_x \partial_y {\widetilde J}_{xy}$, $\partial Q_y / \partial \tau = \int dx \partial_x \partial_y {\widetilde J}_{xy}$, where the additional contribution from the fermion also vanishes because it is a total derivative of the fermion bilinear. Thus, $Q_x$ and $Q_y$ remain conserved so that the subsystem symmetry of the Bose metal is intact for the non-Fermi liquid fixed point.

In the model with multiple low-energy excitations, where the quadratic-band-touching fermions are located at multiple places in the momentum space, we showed that the nodal-line excitations can also influence the large momentum (in one of the two momentum directions) scattering processes between different areas in momentum space. While the contribution from the inter-area scattering process is irrelevant at the $T=0$ fixed point, the non-analytic $T^{3/2}$ correction from such processes may appear in the fermion self-energy, $\text{Im}\Sigma\sim T+ T^{3/2}$, at finite temperature, before one reaches the low temperature scaling regime. For example, the behavior, $\partial^{2}\text{Im}\Sigma/\partial T^{2}\propto T^{-1/2}$, could be seen in the ARPES measurement \cite{RevModPhys.75.473}. 

In the current work, we investigated the interaction between the quadratic-band-touching fermions and the Bose metal. %In particular, 
Here, we may consider cold atom systems with bosons and fermions located at the vertices and the centers of plaquettes for a possible realization of such systems \cite{Dai2017}. The details as to how to engineer the relevant interaction can be found in the Supplementary Materials \cite{SM}.
We focus on the case where there is no renormalization of the nodal-line spectrum of the Bose metal. It would be interesting to investigate the interaction between fermions with a Fermi surface and the Bose metal. Here one may expect that the low energy excitations of the Bose metal would be strongly influenced by the particle-hole excitations of the Fermi sea. This would be an interesting subject of future study.

\emph{Acknowledgement}: We thank Sung-Sik Lee for helpful discussions. 
%This work was supported by the NSERC of Canada and the Center for Quantum Materials at the University of Toronto. 
This work was supported by the NSERC of Canada Grant No.~RGPIN-2017-03774 and the Center for Quantum Materials at the University of Toronto. Y.B.K. is also supported by the Simons Fellowship from the Simons Foundation and the Guggenheim Fellowship from the John Simon Guggenheim Memorial Foundation.

%\bibliographystyle{apsrev4-1}
%\bibliography{ref}
%

\newpage
\onecolumngrid
\clearpage
\begin{center}
\textbf{\large Supplemental Material for ``Non-Fermi liquid induced by the Bose metal with protected sub-system symmetries''}
\end{center}
\begin{center}
{SangEun Han and Yong Baek Kim}\\
\emph{Department of Physics, University of Toronto, Toronto, Ontario M5S 1A7, Canada}\\
\end{center}
\setcounter{equation}{0}
\setcounter{figure}{0}
\setcounter{table}{0}
\setcounter{page}{1}
\setcounter{section}{0}
\setcounter{subsection}{0}

\makeatletter
\renewcommand{\thesection}{\arabic{section}}
\renewcommand{\thesubsection}{\thesection.\arabic{subsection}}
\renewcommand{\thesubsubsection}{\thesubsection.\arabic{subsubsection}}
\renewcommand{\theequation}{S\arabic{equation}}
\renewcommand{\thefigure}{S\arabic{figure}}

\section{Tight-binding Hamiltonian for $p$-orbitals}
Here, we introduce the tight-binding Hamiltonian of the fermions with $p$-orbitals on the square lattice. First, the nearest neighbor hopping Hamiltonian is given by
\begin{align*}
H_{\text{NN}}
=&\sum_{i}t_{\sigma}(c_{x,i+\hat{x}}^{\dagger}c_{x,i}+c_{y,i+\hat{y}}^{\dagger}c_{y,i}+\text{h.c.})+\sum_{i}t_{\pi}(c_{x,i+\hat{y}}^{\dagger}c_{x,i}+c_{y,i+\hat{x}}^{\dagger}c_{y,i}+\text{h.c.})\\
=&\sum_{k}\left(\begin{matrix} \hat{c}_{x,k}^{\dagger}&\hat{c}_{y,k}^{\dagger}\end{matrix}\right)
\left(\begin{matrix} 
2t_{\sigma}\cos(k_{x})+2t_{\pi}\cos(k_{y})&0\\
0&2t_{\sigma}\cos(k_{y})+2t_{\pi}\cos(k_{x})
\end{matrix}\right)
\left(\begin{matrix} \hat{c}_{x,k}\\\hat{c}_{y,k}\end{matrix}\right),
\end{align*}
where h.c. stands for the hermitian conjugate and $c_{(x,y),i}$ and $c_{(x,y),k}$ are the fermion annihilation operators in the position and momentum space for $x$ ($y$) orbital, respectively. $t_{\sigma}$ and $t_{\pi}$ are the hopping parameters for the $\sigma$ and $\pi$ bonds.\\
The next nearest neighbor hopping Hamiltonian is given by
\begin{align*}
H_{\text{NNN}}
=&\sum_{i}\sum_{\alpha=x,y}\tilde{t}(c_{\alpha,i+\hat{x}+\hat{y}}^{\dagger}c_{\alpha,i}+c_{\alpha,i-\hat{x}+\hat{y}}^{\dagger}c_{\alpha,i}+c_{\alpha,i+\hat{x}-\hat{y}}^{\dagger}c_{\alpha,i}+c_{\alpha,i-\hat{x}-\hat{y}}^{\dagger}c_{\alpha,i}+\text{h.c})\\
&+\sum_{i}\sum_{\alpha\neq\beta}\tilde{t}(c_{\alpha,i+\hat{x}+\hat{y}}^{\dagger}c_{\beta,i}-c_{\alpha,i-\hat{x}+\hat{y}}^{\dagger}c_{\beta,i}-c_{\alpha,i+\hat{x}-\hat{y}}^{\dagger}c_{\beta,i}+c_{\alpha,i-\hat{x}-\hat{y}}^{\dagger}c_{\beta,i}+\text{h.c})\\
=&\sum_{k}\left(\begin{matrix} \hat{c}_{x,k}^{\dagger}&\hat{c}_{y,k}^{\dagger}\end{matrix}\right)
\left(\begin{matrix} 
4\tilde{t}\cos(k_{x})\cos(k_{y})&-4\tilde{t}'\sin(k_{x})\sin(k_{y})\\
-4\tilde{t}'\sin(k_{x})\sin(k_{y})&4\tilde{t}\cos(k_{x})\cos(k_{y})
\end{matrix}\right)
\left(\begin{matrix} \hat{c}_{x,k}\\\hat{c}_{y,k}\end{matrix}\right),
\end{align*}
where $\tilde{t}$ and $\tilde{t}'$ are the hopping parameters for the next nearest neighbor hopping. The total tight-binding Hamiltonian is
\begin{align*}
H_{\text{tb}}
=&\sum_{k}\left(\begin{matrix} \hat{c}_{x,k}^{\dagger}&\hat{c}_{y,k}^{\dagger}\end{matrix}\right)
\mathcal{H}(\mathbf{k})
\left(\begin{matrix} \hat{c}_{x,k}\\\hat{c}_{y,k}\end{matrix}\right),
\end{align*}
where
\begin{align*}
\mathcal{H}(\mathbf{k})=\left(\begin{matrix} 
2t_{\sigma}\cos(k_{x})+2t_{\pi}\cos(k_{y})+4\tilde{t}\cos(k_{x})\cos(k_{y})&-4\tilde{t}'\sin(k_{x})\sin(k_{y})\\
-4\tilde{t}'\sin(k_{x})\sin(k_{y})&2t_{\sigma}\cos(k_{y})+2t_{\pi}\cos(k_{x})+4\tilde{t}\cos(k_{x})\cos(k_{y})
\end{matrix}\right).
\end{align*}

Then, near $\mathbf{k}=(0,0)$,
\begin{align*}
\mathcal{H}
=&2(t_{\sigma}+t_{\pi}+2\tilde{t})\sigma_{0}+
\left(\begin{matrix} 
-(t_{\sigma}+2\tilde{t})k_{x}^{2} -(t_{\pi}+2\tilde{t})k_{y}^{2}   &-4\tilde{t}'k_{x}k_{y}\\
-4\tilde{t}'k_{x}k_{y}&-(t_{\pi}+2\tilde{t})k_{x}^{2} -(t_{\sigma}+2\tilde{t})k_{y}^{2}
\end{matrix}\right)\\
=&E_{\Gamma}\sigma_{0}+
\left(A_{1}(k_{x}^{2}+k_{y}^{2})\sigma_{0}+A_{x} (2k_{x}k_{y})\sigma_{x}+A_{z}(k_{x}^{2}-k_{y}^{2})\sigma_{z}\right)
\end{align*}
where $E_{\Gamma}\equiv 2(t_{\sigma}+t_{\pi}+2\tilde{t})$, $A_{0}=-(t_{\sigma}+t_{\pi}+4\tilde{t})/2$, $A_{x}=-2\tilde{t}'$, and $A_{z}=(t_{\pi}-t_{\sigma})a^{2}/2$

\section{Interaction between boson and fermion}
\begin{figure}
\includegraphics{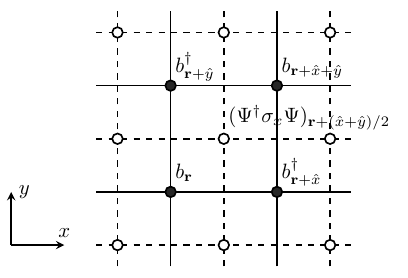}
\caption{The depiction of the interaction between the bosons and fermions on the square lattice. The bosons and fermions reside on the vertex (black dots, $\mathbf{r}$) and the center of the each plaquette (white dots, $\mathbf{r}+(\hat{x}+\hat{y})/2$) in the square lattice, respectively. The solid and dashed lines stand for the square lattices on which the bosons and fermions live, respectively. Here, the ring-exchange interaction of the bosons, $b_{\mathbf{r}}b_{\mathbf{r}+\hat{x}}^{\dagger}b_{\mathbf{r}+\hat{x}+\hat{y}}b_{\mathbf{r}+\hat{y}}^{\dagger}$, couples to the fermion density, $(\Psi^{\dagger}\sigma_{x}\Psi)_{\mathbf{r}+(\hat{x}+\hat{y})/2}$.}\label{figS:bf_lattice}
\end{figure}

Here we discuss how to obtain the interaction term in Eq.~\ref{eq:int} in the lattice model. Considering the fermions and bosons on the two-dimensional square lattice, we may get the symmetry-allowed interaction between bosons and fermions as follows,
\begin{align}
H_{\text{int}}=&\;\frac{g}{2}\sum_{\mathbf{r}}i[b_{\mathbf{r}}b_{\mathbf{r}+\hat{x}}^{\dagger}b_{\mathbf{r}+\hat{x}+\hat{y}}b_{\mathbf{r}+\hat{y}}^{\dagger}\Psi^{\dagger}_{\mathbf{r}+(\hat{x}+\hat{y})/2}\sigma_{x}\Psi_{\mathbf{r}+(\hat{x}+\hat{y})/2}-\text{h.c.}]\label{eq:lattice_int}\notag\\
\notag\sim&\; \frac{g}{2}\sum_{\mathbf{r}}i[e^{-i(\phi_{\mathbf{r}}-\phi_{\mathbf{r}+\hat{x}}-\phi_{\mathbf{r}+\hat{y}}+\phi_{\mathbf{r}+\hat{x}+\hat{y}})}\Psi^{\dagger}_{\mathbf{r}+(\hat{x}+\hat{y})/2}\sigma_{x}\Psi_{\mathbf{r}+(\hat{x}+\hat{y})/2}-\text{h.c.}]\\
\notag=&\;g \sum_{\mathbf{r}}\sin(\Delta_{xy}\phi_{\mathbf{r}})\Psi^{\dagger}_{\mathbf{r}+(\hat{x}+\hat{y})/2}\sigma_{x}\Psi_{\mathbf{r}+(\hat{x}+\hat{y})/2}\\
\notag\approx &\;g\sum_{\mathbf{r}}(\partial_{x}\partial_{y}\phi)(\Psi^{\dagger}\sigma_{x}\Psi),
\end{align}
where bosons and fermions reside on the vertex ($\mathbf{r}$) and the center of each plaquette ($\mathbf{r}+(\hat{x}+\hat{y})/2$) in the square lattice (see Fig.~\ref{figS:bf_lattice}). Here, $b_{\mathbf{r}}^{\dagger}\sim e^{i\phi_{\mathbf{r}}}$ and $b_{\mathbf{r}}\sim e^{-i\phi_{\mathbf{r}}}$. 
As a result, we can get the coupling term in Eq.~\ref{eq:int}. % from Eq.~\ref{eq:lattice_int}.
Note that there is another choice for the interaction between bosons and fermions, $\cos(\Delta_{xy}\phi_{\mathbf{r}})(\Psi^{\dagger}\Psi)_{\mathbf{r}+(\hat{x}+\hat{y})/2}$, but its leading nontrivial contribution in continuum limit is $(\partial_{x}\partial_{y}\phi)^{2}(\Psi^{\dagger}\Psi)$ and it is irrelevant.

\section{Definition of $F_{i}$ functions}
The functions $F_{\omega,0,x,z}$ in Eq.~\ref{eq:deltaSigma} and vertex correction 
 in the main text are defined as
\begin{align}
F_{\omega}(a_{0},a_{x},a_{z})
=&\int_{-\infty}^{\infty}\frac{d\omega}{2\pi}\int_{0}^{\pi/4}d\theta\;
\frac{\omega^{2}\mathbb{S}(\theta)(\omega^{2}+a_{0}^{2}+a_{x}^{2}\mathbb{S}(\theta)+a_{z}^{2}\mathbb{C}(\theta))}{(\omega^{2}+\mathbb{S}(\theta)/4)^{2}(\omega^{4}+2(a_{0}^{2}+a_{x}^{2}\mathbb{S}(\theta)+a_{z}^{2}\mathbb{C}(\theta))+(a_{0}^{2}-a_{x}^{2}\mathbb{S}(\theta)-a_{z}^{2}\mathbb{C}(\theta))^{2})}
\\
%%%%
F_{0}(a_{0},a_{x},a_{z})
=&\int_{-\infty}^{\infty}\frac{d\omega}{2\pi}\int_{0}^{\pi/4}d\theta\;
\frac{\omega^{2}(-\omega^{2}+3\mathbb{S}(\theta)/4)(\omega^{2}+a_{0}^{2}-a_{x}^{2}\mathbb{S}(\theta)-a_{z}^{2}\mathbb{C}(\theta))}{(\omega^{2}+\mathbb{S}(\theta)/4)^{3}(\omega^{4}+2(a_{0}^{2}+a_{x}^{2}\mathbb{S}(\theta)+a_{z}^{2}\mathbb{C}(\theta))+(a_{0}^{2}-a_{x}^{2}\mathbb{S}(\theta)-a_{z}^{2}\mathbb{C}(\theta))^{2})},
\\
%%%
F_{x}(a_{0},a_{x},a_{z})
=&\int_{-\infty}^{\infty}\frac{d\omega}{2\pi}\int_{0}^{\pi/4}d\theta\;
\frac{4\omega^{2}\mathbb{S}(\theta)(-\omega^{2}+\mathbb{S}(\theta)/4)(\omega^{2}-a_{0}^{2}+a_{x}^{2}\mathbb{S}(\theta)+a_{z}^{2}\mathbb{C}(\theta))}{(\omega^{2}+\mathbb{S}(\theta)/4)^{3}(\omega^{4}+2(a_{0}^{2}+a_{x}^{2}\mathbb{S}(\theta)+a_{z}^{2}\mathbb{C}(\theta))+(a_{0}^{2}-a_{x}^{2}\mathbb{S}(\theta)-a_{z}^{2}\mathbb{C}(\theta))^{2})}
,\\
%%%
F_{z}(a_{0},a_{x},a_{z})
=&\int_{-\infty}^{\infty}\frac{d\omega}{2\pi}\int_{0}^{\pi/4}d\theta\;
\frac{\omega^{2}\mathbb{C}(\theta)(-\omega^{2}+3\mathbb{S}(\theta)/4)(\omega^{2}-a_{0}^{2}+a_{x}^{2}\mathbb{S}(\theta)+a_{z}^{2}\mathbb{C}(\theta))}{(\omega^{2}+\mathbb{S}(\theta)/4)^{3}(\omega^{4}+2(a_{0}^{2}+a_{x}^{2}\mathbb{S}(\theta)+a_{z}^{2}\mathbb{C}(\theta))+(a_{0}^{2}-a_{x}^{2}\mathbb{S}(\theta)-a_{z}^{2}\mathbb{C}(\theta))^{2})},
\\
%%%
F_{g}(a_{0},a_{x},a_{z})
=&\int_{-\infty}^{\infty}\frac{d\omega}{2\pi}\int_{0}^{\pi/4}d\theta\;
\frac{\mathbb{S}(\theta)/2
(a_{0}^{6}+a_{0}^{4}(\omega^{2}-a_{x}^{2}\mathbb{S}(\theta)-3a_{z}^{2}\mathbb{C}(\theta))-(\omega^{2}-a_{x}^{2}\mathbb{S}(\theta)+a_{z}^{2}\mathbb{C}(\theta))(\omega^{2}+a_{x}^{2}\mathbb{S}(\theta)+a_{z}^{2}\mathbb{C}(\theta))^{2}
)}{(\omega^{2}+\mathbb{S}(\theta)/4)(\omega^{4}+2(a_{0}^{2}+a_{x}^{2}\mathbb{S}(\theta)+a_{z}^{2}\mathbb{C}(\theta))+(a_{0}^{2}-a_{x}^{2}\mathbb{S}(\theta)-a_{z}^{2}\mathbb{C}(\theta))^{2})^{2}}\notag\\
&-\int_{-\infty}^{\infty}\frac{d\omega}{2\pi}\int_{0}^{\pi/4}d\theta\;
\frac{a_{0}^{2}\mathbb{S}(\theta)/2
(\omega^{4}+2\omega^{2}(5a_{x}^{2}\mathbb{S}(\theta)-a_{z}^{2}\mathbb{C}(\theta))+(a_{x}^{2}\mathbb{S}(\theta)+a_{z}^{2}\mathbb{C}(\theta))(a_{x}^{2}\mathbb{S}(\theta)-3a_{z}^{2}\mathbb{C}(\theta)))
}{(\omega^{2}+\mathbb{S}(\theta)/4)(\omega^{4}+2(a_{0}^{2}+a_{x}^{2}\mathbb{S}(\theta)+a_{z}^{2}\mathbb{C}(\theta))+(a_{0}^{2}-a_{x}^{2}\mathbb{S}(\theta)-a_{z}^{2}\mathbb{C}(\theta))^{2})^{2}},
\end{align}
where $\mathbb{S}(\theta)=\sin^{2}(2\theta)$ and $\mathbb{C}(\theta)=\cos^{2}(2\theta)$. Here, $F_{\omega,0,g}$ are positive, but $F_{z}$ is negative for $a_{0}>a_{x},a_{z}$. Note that the difference between $F_{\omega}$ and $F_{g}$ is given by
\begin{align}
F_{\omega}-F_{g}
=&\int_{0}^{2\pi}d\theta\;\frac{2a_{z}^{3}\cos^{2}(2\theta)|\sin(2\theta)|}{((2a_{0}+|\sin(2\theta)|)^{2}-4(a_{1}^{2}\sin^{2}(2\theta)+a_{z}^{2}\cos^{2}(2\theta))^{2}}\geq0,\label{eq:FoFg}
\end{align}
in the parameter region $a_{0}>a_{x,z}$, which we consider in the main text. The equality holds for $a_{z}=0$.

When $a_{z}\rightarrow0$, $F_{\omega,0,x,g}$ can be written as
\begin{align}
F_{\omega}(a_{0},a_{x},0)\equiv&\frac{-1+4a_{0}^{2}-4a_{x}^{2}}{((1-2a_{0})^{2}-4a_{x}^{2})((1+2a_{0})^{2}-4a_{x}^{2})}\\
&+\frac{(1-2a_{x})\tan^{-1}\left[\frac{1-2a_{0}-2a_{x}}{\sqrt{4a_{0}^{2}-(1-2a_{x})^{2}}}\right]}{(4a_{0}^{2}-(1-2a_{x})^{2})^{3/2}}+\frac{(1+2a_{x})\tan^{-1}\left[\frac{1-2a_{0}+2a_{x}}{\sqrt{4a_{0}^{2}-(1+2a_{x})^{2}}}\right]}{(4a_{0}^{2}-(1+2a_{x})^{2})^{3/2}},\\
%%%%
F_{0}(a_{0},a_{x},0)\equiv&\frac{-256a_{0}^{6}+(1-4a_{x}^{2})^{2}(1-12a_{x}^{2})+16a_{0}^{4}(9+20a_{x}^{2})+8a_{0}^{2}(-3-24a_{x}^{2}+16a_{x}^{4})}{4a_{0}^{2}(16a_{0}^{4}+(1-4a_{x}^{2})^{2}-8a_{0}^{2}(1+4a_{x}^{2}))^{2}}\\
&-\frac{(1+8a_{0}^{2}+2a_{x}-8a_{x}^{2})\tan^{-1}\left[\frac{1-2a_{0}-2a_{x}}{\sqrt{4a_{0}^{2}-(1-2a_{x})^{2}}}\right]}{(4a_{0}^{2}-(1-2a_{x})^{2})^{5/2}}-\frac{(1+8a_{0}^{2}-2a_{x}-8a_{x}^{2})\tan^{-1}\left[\frac{1-2a_{0}+2a_{x}}{\sqrt{4a_{0}^{2}-(1+2a_{x})^{2}}}\right]}{(4a_{0}^{2}-(1-2a_{x})^{2})^{5/2}},\\
F_{x}(a_{0},a_{x},0)\equiv&\frac{6(-1+48a_{0}^{4}-8a_{x}^{2}+48a_{x}^{4}-8a_{0}^{2}(1+12a_{x}^{2}))}{(16a_{0}^{4}+(1-4a_{x}^{2})^{2}-8a_{0}^{2}(1+4a_{x}^{2}))^{2}}\\
&-\frac{2(8a_{0}^{4}+a_{x}(1+2a_{x})(1-2a_{x})^{2}+4a_{0}^{2}(1+a_{x}-4a_{x}^{2}))\tan^{-1}\left[\frac{1-2a_{0}-2a_{x}}{\sqrt{4a_{0}^{2}-(1-2a_{x})^{2}}}\right]}{a_{x}(4a_{0}^{2}-(1-2a_{x})^{2})^{5/2}}\\
&+\frac{2(8a_{0}^{4}-a_{x}(1-2a_{x})(1+2a_{x})^{2}+4a_{0}^{2}(1-a_{x}-4a_{x}^{2}))\tan^{-1}\left[\frac{1-2a_{0}+2a_{x}}{\sqrt{4a_{0}^{2}-(1+2a_{x})^{2}}}\right]}{a_{x}(4a_{0}^{2}-(1+2a_{x})^{2})^{5/2}},\\
F_{g}(a_{0},a_{x},0)=&F_{\omega}(a_{0},a_{x},0).
\end{align}

\section{RG flow equations of parameters}

\begin{figure}
\subfigure[]{
\includegraphics[height=14.1em]{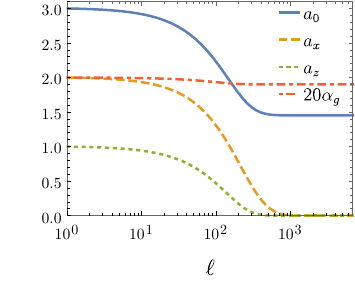}\label{figS:rgflow_a}}
\subfigure[]{
\includegraphics[height=14.1em]{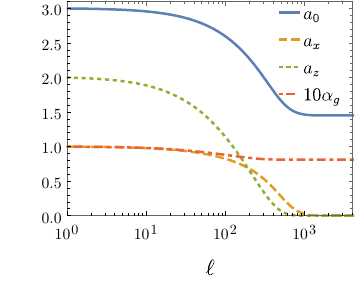}\label{figS:rgflow_b}}
\subfigure[]{
\includegraphics[height=14.1em]{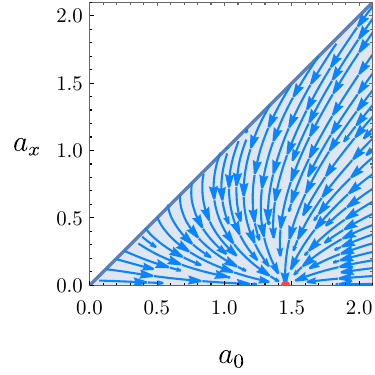}\label{figS:rgflow_c}}
\caption{
(a,b) The RG flows for some initial values of the dimensionless parameters. (a) The initial value $(a_{0},a_{x},a_{z},\alpha_{g})=(3,2,1,0.1)$. It has the fixed point, $(a_{0}^{*},a_{x}^{*},a_{z}^{*},\alpha_{g}^{*})=(1.4518,0,0,0.0951)$. (b) The initial value $(a_{0},a_{x},a_{z},\alpha_{g})=(3,1,2,0.1)$. It has the fixed point, $(a_{0}^{*},a_{x}^{*},a_{z}^{*},\alpha_{g}^{*})=(1.4518,0,0,0.0811)$. 
(c) The RG flow in terms of $a_{0}$ and $a_{x}$ for an arbitrary fixed $\alpha_{g}$ when $a_{z} = 0$. The red point represents the stable fixed point values of $a_{0}$ and $a_{x}$, $(a_{0}^{*},a_{x}^{*})=(1.4518,0)$.}\label{figS:1}
\end{figure}

The detailed expressions of the one-loop corrections in the main text are given by
\begin{align}
\delta \Pi =&
-g^{2}q_{x}^{2}q_{y}^{2}\int_{\partial\Lambda}\text{Tr}[\sigma_{x}G(i\Omega_{m}+i\omega_{n},\mathbf{p}+\mathbf{q})\sigma_{x}G(i\Omega_{m},\mathbf{p})]=0,\\
\delta \Sigma (i\omega_{n},\mathbf{k})
=&\; g^{2}\int_{\partial\Lambda}(p_{x}-k_{x})^{2}(p_{y}-k_{y})^{2}\sigma_{x}G(i\Omega_{m},\mathbf{p})\sigma_{x} D(i\Omega_{m}-i\omega_{n},\mathbf{p}-\mathbf{k})\notag \\
\approx&
-\alpha_{g}\ell[ F_{\omega}(-i\omega_{n})\sigma_{0}+F_{0}A_{0}k^{2}\sigma_{0}+F_{x}(2A_{x}k_{x}k_{y})\sigma_{x}+F_{z}A_{z}(k_{x}^{2}-k_{y}^{2})
],\\
\delta \Gamma_{g}=&\;\frac{g^{2}}{2}\int_{\partial\Lambda}p_{x}^{2}p_{y}^{2}D(i\Omega_{m},\mathbf{p})\text{Tr}[\sigma_{x}G(i\Omega_{m},\mathbf{p})\sigma_{x}G(i\Omega_{m},\mathbf{p})]\notag\\
=&\;\alpha_{g}F_{g}\ell,
\end{align}
where $\delta\Pi$, $\delta\Sigma$, and $\delta\Gamma_{g}$ are the boson self-energy, fermion self-energy, and vertex correction, respectively. $\ell=\ln(\Lambda/\nu)$, $\Lambda$ and $\mu$ are UV and IR cutoffs. The one-loop boson self-energy $\delta \Pi=0$ when $A_{0}>\text{max}(A_{1},A_{3})$, as mentioned in the main text. $F_{i}$'s are functions of $a_{0}$, $a_{x}$, and $a_{z}$, and their definitions are provided in the previous section.
From the one-loop corrections, we obtain the following RG flow equations for the parameters of the model,
\begin{align}
\frac{1}{A_{i}}\frac{dA_{i}}{d\ell}=&(z-2)-\alpha_{g}(F_{\omega}-F_{i}),\\
\frac{1}{u}\frac{du}{d\ell}=&(z-2),\\
\frac{1}{g}\frac{dg}{d\ell}=&\frac{1}{2}(3z-d-4)-\alpha_{g}(F_{\omega}-F_{g}).
\end{align}
Combining the RG flow equations above, 
we obtain the RG flow equations of the dimensionless parameters,
\begin{align}
\frac{1}{a_{i}}\frac{da_{i}}{d\ell}=&-\alpha_{g}(F_{\omega}-F_{i}),\label{eqS:RG1}\\
\frac{1}{\alpha_{g}}\frac{d\alpha_{g}}{d\ell}=&(2-d)-2\alpha_{g}(F_{\omega}-F_{g}).\label{eqS:RG2}
\end{align}
Note that Eq.~\ref{eqS:RG2} is negative for $d=2$ since $-(F_{\omega}-F_{g})$ is negative in the parameter region $a_{0}>a_{x,z}$ (see Eq.~\ref{eq:FoFg}).
As mentioned in the main text, the RG flow equations Eq.~\ref{eqS:RG1} and \ref{eqS:RG2} lead to a line of stable fixed points, $(a_{0}^{*},a_{x}^{*},a_{z}^{*},\alpha_{g}^{*})=(1.4518,0,0,\alpha_{g}^{*})$, where $\alpha_{g}^{*}$ depends on the initial values of the dimensionless parameters.
Near the fixed point, $(a_{0},a_{x},a_{z},\alpha_{g})=(a_{0}^{*}+\delta a_{0},\delta a_{x},\delta a_{z},\alpha_{g}^{*}+\delta \alpha_{g})$, the flow equations of $\delta a_{0},\delta a_{x},\delta a_{z},\delta \alpha_{g}$ are
\begin{align*}
\frac{d \delta a_{0}}{d\ell}=& -0.07073\delta\alpha_{g}\delta a_{0},\quad
\frac{d \delta a_{x}}{d\ell}=-0.06707\delta\alpha_{g}\delta a_{x},\quad
\frac{d \delta a_{z}}{d\ell}=-0.1488\delta\alpha_{g}\delta a_{z},\quad
\frac{d\alpha_{g}}{d\ell}=0.
\end{align*}
For example, the RG flows for several initial values of the parameters are shown in Fig.~S\ref{figS:rgflow_a} and S\ref{figS:rgflow_b}. The fixed point values of the parameters in Fig.~S\ref{figS:rgflow_a} and S\ref{figS:rgflow_b} belong to the line of stable fixed points.
As $a_{z}\rightarrow0$, $a_{0}$ and $a_{x}$ approach the fixed point values, $(a_{0}^{*},a_{x}^{*})=(1.4518,0)$, represented by the red point in Fig.~S\ref{figS:rgflow_c}.

\section{Anomalous dimension by Field-theoretical RG}
Here, using the field-theoretical RG, we find the anomalous dimension of the fermions from the one-loop fermion self-energy. The Callan-Symanzik equation for the fermion propagator is given by
\begin{align}
\left(\Lambda\frac{\partial}{\partial\Lambda}+\beta_{g}\frac{\partial}{\partial g}+\sum_{i}\beta_{A_{i}}\frac{\partial}{\partial A_{i}}+\beta_{u}\frac{\partial}{\partial u}+\eta_{f}\right)G=0,\label{eq:callan_symanzik}
\end{align}
where $\eta_{f}$ is the anomalous dimension of the fermions, and $\Lambda$ is the UV cutoff. $\beta_{g}$, $\beta_{A_{i}}$, and $\beta_{u}$ are the beta equations of $g$, $A_{i}$, and $u$, and they are defined as
\begin{align}
\beta_{g}=\Lambda\frac{\partial g}{\partial\Lambda},\quad\beta_{A_{i}}=\Lambda\frac{\partial A_{i}}{\partial\Lambda},\quad\beta_{u}=\Lambda\frac{\partial u}{\partial\Lambda},\quad\eta_{f}=-\Lambda\frac{\partial \ln Z_{\psi}^{-1}}{\partial\Lambda},
\end{align}
respectively, and $Z_{\psi}$ is wavefucntion the renormalization constant of the fermions. Assuming that the parameters are near the fixed point values, we have $\beta_{A_{i}}=\beta_{u}=\beta_{g}=0$. Then, Eq.~\ref{eq:callan_symanzik} can be written as $\eta_{f}=-\frac{\Lambda}{G}\frac{\partial G}{\partial\Lambda}.$
The one-loop fermion self-energy has the form, $\Sigma(i\omega_{n})\approx-i\omega_{n}\alpha_{g}C_{\text{log}}\ln\Big(\frac{|\omega_{n}|}{u\Lambda^{2}}\Big)+\text{regular}+\cdots$, as shown in the main text (where $C_{\text{log}}>0$). From this, we obtain $\frac{\Lambda}{G}\frac{\partial G}{\partial\Lambda}=-2\alpha_{g}C_{\text{log}}$. Hence the anomalous dimension of the fermions is $\eta_{f}=2\alpha_{g}C_{\text{log}}$.
For example, when we have the fixed point value $(a_{0},a_{x},a_{z})=(1.4518,0,0)$, using $C_{\text{log}}=0.03719$, we obtain $\eta_{f}=0.07438\alpha_{g}$. Note that along the line of stable fixed points, $C_{\text{log}}$ does not change.

From the perturbative Wilsonian RG, we can also obtain the anomalous dimension of the fermions, $\eta_{f}=-\Lambda\frac{\partial \ln Z_{\psi}^{-1}}{\partial\Lambda}=\alpha_{g}F_{\omega}$, where $Z_{\psi}=1+\alpha_{g}F_{\omega}\ln({\Lambda}/{\mu})$ and $F_{\omega}(1.4518,0,0)=C_{f}=0.07439$ at the fixed point. Then, we can check that the anomalous dimension of the fermions from the field-theoretical RG and Wilsonian RG is consistent, and $2C_{\text{log}}\approx C_{f}$.

\section{One-loop fermion self-energies from small and large momentum transfer}

The one-loop fermion self-energy for the small momentum transfer within the same area is
\begin{align}
\Sigma_{\text{small}}&(i\omega_{m})\notag\\
=&g^{2}\int_{\Lambda} G(i\Omega_{m}+i\omega_{n},\mathbf{p})D(i\Omega_{m},\mathbf{p})\mathcal{F}(\mathbf{p})^{2}\notag\\
=&g^{2}\int_{\Lambda}\frac{dp_{x}dp_{y}}{(2\pi)^{2}}\int_{-u\Lambda^{2}}^{u\Lambda^{2}}\frac{d\Omega_{m}}{2\pi}\frac{1}{-i(\Omega_{m}+\omega_{n})+A_{0}p^{2}}\frac{p_{x}^{2}p_{y}^{2}}{\Omega_{m}^{2}+u^{2}p_{x}^{2}p_{y}^{2}}\notag\\
=&\frac{g^{2}\Lambda^{2}}{8\pi^{2}u^{2}}\int_{0}^{\Lambda}\frac{2pdp}{\Lambda^{2}}\int_{-u\Lambda^{2}}^{u\Lambda^{2}}\frac{d\Omega_{m}}{u\Lambda^{2}}\int_{0}^{2\pi}\frac{d\theta}{2\pi}\frac{1}{-i(\Omega_{m}+\omega_{n})/u\Lambda^{2}+(A_{0}/u)(p/\Lambda)^{2}}\frac{(p/\Lambda)^{4}\sin^{2}(2\theta)/4}{(\Omega_{m}/u\Lambda^{2})^{2}+(p/\Lambda)^{4}\sin^{2}(2\theta)/4}\notag\\
=&\alpha_{g}\frac{u\Lambda^{2}}{8}\int_{0}^{1}dx\int_{-1}^{1}d\Omega\int_{0}^{2\pi}\frac{d\theta}{2\pi}\frac{1}{-i(\Omega+(\omega_{n}/u\Lambda^{2}))+a_{0}x}\frac{x^{2}\sin^{2}(2\theta)/4}{\Omega^{2}+x^{2}\sin^{2}(2\theta)/4}\notag\\
=&\alpha_{g}\frac{u\Lambda^{2}}{8}\int_{0}^{1}dx\int_{-1}^{1}d\Omega\frac{1}{-i(\Omega+(\omega_{n}/u\Lambda^{2}))+a_{0}x}\left(1-\frac{|\Omega|}{\sqrt{\Omega^{2}+x^{2}/4}}\right)\label{eqS:smalleq}\\
\approx&\alpha_{g}\left(\mathcal{C}_{\mathcal{R},1}^{\text{small}}|\omega_{n}|+\mathcal{C}_{\mathcal{R},2}^{\text{small}}\frac{\omega_{n}^{2}}{u\Lambda^{2}}+\mathcal{C}_{\mathcal{I},1}^{\text{small}}i\omega_{n}+\mathcal{C}_{\mathcal{I},\text{log}}^{\text{small}}i\omega_{n}\ln\Big(\frac{|\omega_{n}|}{u\Lambda^{2}}\Big)\right),\label{eqS:small}
\end{align}
where $G(i\Omega_{m},\mathbf{p})=(-i\Omega_{n}+A_{0}p^{2})^{-1}$, $D(i\Omega_{m},\mathbf{p})=(\Omega_{m}^{2}+u^{2}p_{x}^{2}p_{y}^{2})^{-1}$, and $\mathcal{F}(\mathbf{p})=p_{x}p_{y}$ are the propagators of the fermions and bosons, and the form factor of the interaction, respectively, $x\equiv (p/\Lambda)^{2}$ and $\Omega\equiv \Omega_{m}/u\Lambda^{2}$. The coefficients $C_{i}^{\text{small}}$'s depend on $a_{0}$. The logarithmic correction comes from the last term in Eq.~\ref{eqS:smalleq}.\\
The one-loop fermion self-energy for the large momentum transfer between different areas in momentum space is
\begin{align}
\Sigma_{\text{large}}&(i\omega_{m})\notag\\
=&g^{2}\int_{\Lambda} G(i\Omega_{m}+i\omega_{n},\mathbf{p})D(i\Omega_{m},\mathbf{p}-2k_{0}\hat{x})\mathcal{F}(\mathbf{p}-2k_{0}\hat{x})\notag\\
=&g^{2}\int_{\Lambda}\frac{dp_{x}dp_{y}}{(2\pi)^{2}}\int_{-u\Lambda^{2}}^{u\Lambda^{2}}\frac{d\Omega_{m}}{2\pi}\frac{1}{-i(\Omega_{m}+\omega_{n})+A_{0}p^{2}}\frac{4\sin^{2}(k_{0})p_{y}^{2}}{\Omega_{m}^{2}+4u^{2}\sin^{2}(k_{0})p_{y}^{2}}\notag\\
=&\frac{g^{2}}{8\pi^{2}u^{3}}\frac{u\Lambda^{2}}{8}\int_{0}^{\Lambda}\frac{2pdp}{\Lambda^{2}}\int_{-u\Lambda^{2}}^{u\Lambda^{2}}\frac{d\Omega_{m}}{u\Lambda^{2}}\int_{0}^{2\pi}\frac{d\theta}{2\pi}\frac{1}{-i(\Omega_{m}+\omega_{n})/u\Lambda^{2}+(A_{0}/u)(p/\Lambda)^{2}}\frac{4\Lambda^{-2}\sin^{2}(k_{0})\sin^{2}\theta (p/\Lambda)^{2}}{(\Omega_{m}/u\Lambda^{2})^{2}+4\Lambda^{-2}\sin^{2}(k_{0})\sin^{2}\theta(p/\Lambda)^{2}}\notag\\
=&\alpha_{g}\frac{u\Lambda^{2}}{8}\int_{0}^{1}dx\int_{-1}^{1}d\Omega\int_{0}^{2\pi}\frac{d\theta}{2\pi}\frac{1}{-i(\Omega+(\omega_{n}/u\Lambda^{2}))+a_{0}x}\frac{\xi x\sin^{2}\theta}{\Omega^{2}+\xi x\sin^{2}\theta}\notag\\
=&\alpha_{g}\frac{u\Lambda^{2}}{8}\int_{0}^{1}dx\int_{-1}^{1}d\Omega\frac{1}{-i(\Omega+(\omega_{n}/u\Lambda^{2}))+a_{0}x}\left(1-\frac{|\Omega|}{\sqrt{\Omega^{2}+\xi x}}\right)\label{eqS:largeeq}\\
\approx& \alpha_{g}\left(\mathcal{C}_{\mathcal{R},3/2}^{\text{large}}\frac{|\omega_{n}|^{3/2}}{(u\Lambda^{2})^{1/2}}+\mathcal{C}_{\mathcal{R},2}^{\text{large}}\frac{\omega_{n}^{2}}{u\Lambda^{2}}+\mathcal{C}_{\mathcal{I},1}^{\text{large}}i\omega_{n}+\mathcal{C}_{\mathcal{I},3/2}^{\text{large}}\frac{i\text{sgn}(\omega_{n})|\omega_{n}|^{3/2}}{(u\Lambda^{2})^{1/2}}\right),\label{eqS:large}
\end{align}
where $G(i\Omega_{m},\mathbf{p})=(-i\Omega_{n}+A_{0}p^{2})^{-1}$, $D(i\Omega_{m},\mathbf{p}-2k_{0}\hat{x})=(\Omega_{m}^{2}+4u^{2}\sin^{2}(k_{0})p_{y}^{2})^{-1}$, and $\mathcal{F}(\mathbf{p}-2k_{0}\hat{x})=2\sin(k_{0})p_{y}$ are the propagators of the fermions and bosons, and the form factor of the interaction, respectively, and $\xi\equiv 4\Lambda^{-2}\sin^{2}(k_{0})$. 
The coefficients $C_{i}^{\text{large}}$'s depend on $a_{0}$ and $\xi$. The non-analytic $|\omega_{n}|^{3/2}$ comes form the last term in Eq.~\ref{eqS:largeeq}. Note that in the real parts of $\Sigma_{\text{small}}$ and $\Sigma_{\text{large}}$ have constant terms which depend on the cutoff, but we drop those.

\begin{figure}
\subfigure[]{
\includegraphics[width=0.40\linewidth]{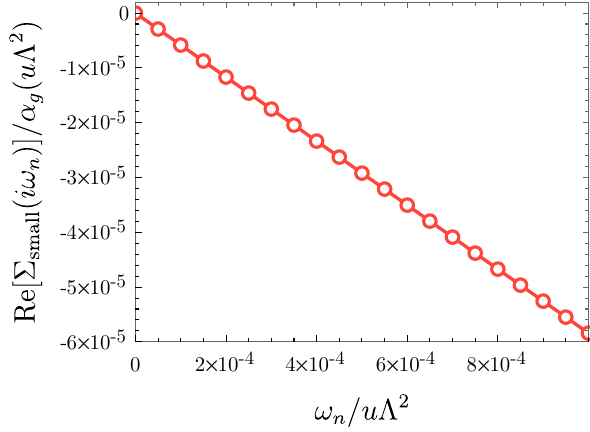}}\;\;
\subfigure[]{
\includegraphics[width=0.40\linewidth]{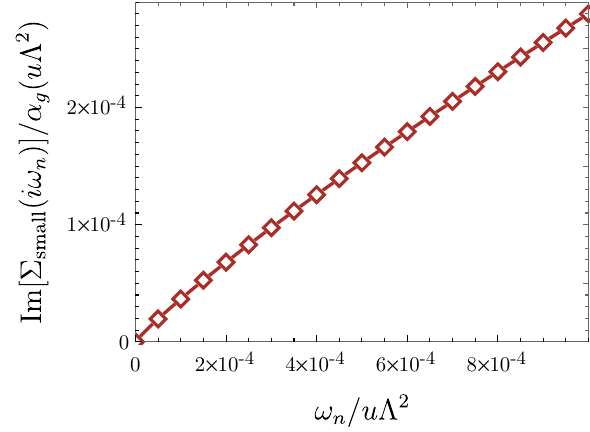}}\\
\subfigure[]{
\includegraphics[width=0.40\linewidth]{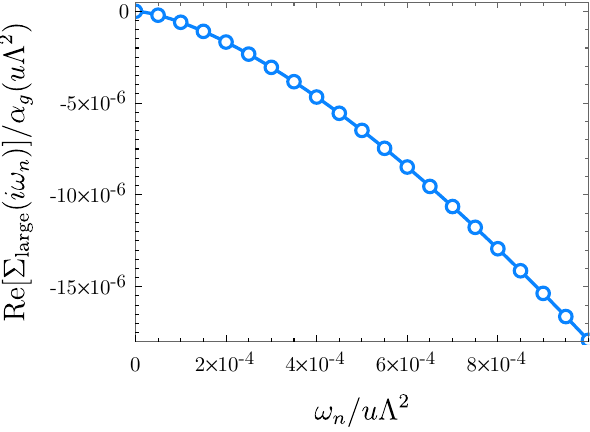}}\;\;
\subfigure[]{
\includegraphics[width=0.40\linewidth]{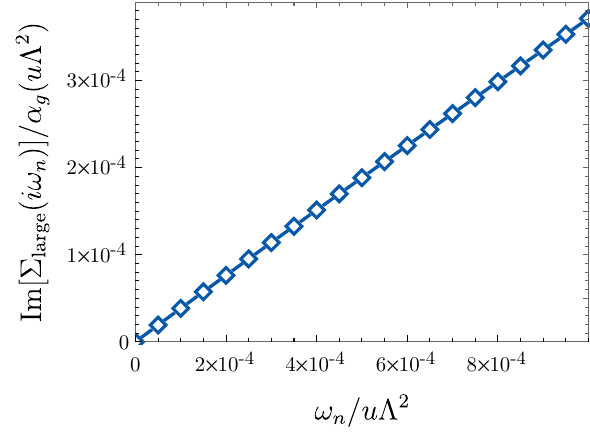}}
\caption{Comparison between the numerical calculation and the functional forms for $\Sigma_{\text{small}}(i\omega_{n})$ and $\Sigma_{\text{large}}(i\omega_{n})$ when $a_{0}=1.4518$ and $\xi=1$. (a), (b) The real and imaginary parts of $\Sigma_{\text{small}}(i\omega_{n})$. The red circles and diamonds are the numerical values and solid lines stand for the fitting of the functional forms. (c), (d) The real and imaginary parts of $\Sigma_{\text{large}}(i\omega_{n})$. The blue circles and diamonds are the numerical values and solid lines stand for the fitting of the functional forms. (a-d) are well-fitted by Eq.~\ref{eqS:small} and \ref{eqS:large}.
}\label{figS:numfit}
\end{figure}

The comparison between the numerical values and fitting functions of $\Sigma_{\text{small}}(i\omega_{n})$ and $\Sigma_{\text{large}}(i\omega_{n})$ for $a_{0}=1.4518$ and $\xi=1$ is shown in Fig.~\ref{figS:numfit}. The coefficients are $\mathcal{C}_{\mathcal{R},1}^{\text{small}}=-0.05842$, $\mathcal{C}_{\mathcal{R},2}^{\text{small}}=-0.01951$, $\mathcal{C}_{\mathcal{I},1}^{\text{small}}=0.02305$, and $\mathcal{C}_{\mathcal{I},\text{log}}^{\text{small}}=-0.03719$ for the small momentum transfer, $\mathcal{C}_{\mathcal{R},3/2}^{\text{large}}=-0.6132$, $\mathcal{C}_{\mathcal{R},2}^{\text{large}}=1.466$, $\mathcal{C}_{\mathcal{I},1}^{\text{large}}=0.3905$, $\mathcal{C}_{\mathcal{I},3/2}^{\text{large}}=-0.6110$ for the large momentum transfer, respectively.\\
\indent After analytic continuation, $i\omega_{n}\rightarrow\omega+i\eta$ (assuming $\omega>0$ for simplicity),
\begin{align}
\Sigma_{\text{small}}(\omega)\approx&\;\alpha_{g}\left(C_{\mathcal{R},1}^{\text{small}}\omega+C_{\mathcal{R},\text{log}}^{\text{small}}\omega\ln\left(\frac{\omega}{u\Lambda^{2}}\right)+C_{\mathcal{R},2}^{\text{small}}\frac{\omega^{2}}{u\Lambda^{2}}
+C_{\mathcal{I},1}^{\text{small}}i\omega\right),\\
\Sigma_{\text{large}}(\omega)\approx&\;\alpha_{g}\left(C_{\mathcal{R},1}^{\text{large}}\omega+C_{\mathcal{R},3/2}^{\text{large}}\frac{\omega^{3/2}}{(u\Lambda^{2})^{1/2}}
+C_{\mathcal{R},2}^{\text{large}}\frac{\omega^{2}}{u\Lambda^{2}}
+C_{\mathcal{I},3/2}^{\text{large}}\frac{i\omega^{3/2}}{(u\Lambda^{2})^{1/2}}\right).
\end{align}
The total one-loop fermion self-energy is
\begin{align}
\Sigma_{\text{tot}}(\omega)\approx&\;\alpha_{g}\left(C_{\mathcal{R},1}^{\text{tot}}\omega+C_{\mathcal{R},\text{log}}^{\text{tot}}\omega\ln\Big(\frac{\omega}{u\Lambda^{2}}\Big)+C_{\mathcal{R},3/2}^{\text{tot}}\frac{\omega^{3/2}}{(u\Lambda^{2})^{1/2}}+C_{\mathcal{R},2}^{\text{tot}}\frac{\omega^{2}}{u\Lambda^{2}}+C_{\mathcal{I},1}^{\text{tot}}i\omega+C_{\mathcal{I},3/2}^{\text{tot}}\frac{i\omega^{3/2}}{(u\Lambda^{2})^{1/2}}\right).
\end{align}

\section{Boson self-energy in multiple low-energy model}
Here, we show that both the small and large momentum transfer processes do not renormalize the bosonic excitations in the model with multiple low-energy excitations.
The one-loop boson self-energy for the small momentum transfer is given by
\begin{align*}
\Pi(i\omega_{n},\mathbf{q})=&-g^{2} q_{x}^{2} q_{y}^{2}\int_{\Lambda}\frac{d^{2}p}{(2\pi)^{2}}\int_{-\infty}^{\infty}\frac{d\Omega_{m}}{2\pi}G(i\Omega_{m}+i\omega_{n},\mathbf{p}+\mathbf{q})G(i\Omega_{m},\mathbf{p})\\
=&-g^{2} q_{x}^{2} q_{y}^{2}\int_{\Lambda}\frac{d^{2}p}{(2\pi)^{2}}\int_{-\infty}^{\infty}\frac{d\Omega_{m}}{2\pi}\frac{1}{-i(\Omega_{m}+\omega_{n})+A_{0}(\mathbf{p}+\mathbf{q})^{2}}\frac{1}{-i\Omega_{m}+A_{0}p^{2}}\\
=&0.
\end{align*}
In terms of $\Omega_{m}$, the integrand has poles at $-\omega_{n}-i A_{0}(\mathbf{p}+\mathbf{q})^{2}$ and $-iA_{0}p^{2}$. If we choose to close the $d\Omega_{m}$ contour in the upper half-plane, the integral becomes zero because two poles are in the lower half-plane. Therefore, $\Pi=0$.\\
The one-loop boson self-energy for the large momentum transfer is given by
\begin{align*}
\Pi(i\omega_{n},\mathbf{q}-2k_{0}\hat{x})=&-g^{2} (q_{x}-2k_{0})^{2} q_{y}^{2}\int_{\Lambda}\frac{d^{2}p}{(2\pi)^{2}}\int_{-\infty}^{\infty}\frac{d\Omega_{m}}{2\pi}G(i\Omega_{m}+i\omega_{n},\mathbf{p}+\mathbf{q}-2k_{0}\hat{x})G(i\Omega_{m},\mathbf{p})\\
=&-g^{2} (q_{x}-2k_{0})^{2} q_{y}^{2}\int_{\Lambda}\frac{d^{2}p}{(2\pi)^{2}}\int_{-\infty}^{\infty}\frac{d\Omega_{m}}{2\pi}\frac{1}{-i(\Omega_{m}+\omega_{n})+A_{0}(\mathbf{p}+\mathbf{q}-2k_{0}\hat{x})^{2}}\frac{1}{-i\Omega_{m}+A_{0}p^{2}}\\
=&\;0.
\end{align*}
It also vanishes in the same way that the one-loop boson self-energy for the small momentum transfer does. Therefore, the small and large momentum processes in the model with multiple low-energy excitations do not renormalize the bosonic action of the Bose metal.

\end{document}